\documentstyle[sprocl]{article}

\def\al{\alpha}

\def\be{\begin{equation}}
\def\ee{\end{equation}}
\def\bea{\begin{eqnarray}}
\def\eea{\end{eqnarray}}

\def\al{\alpha}
\def\be{\beta}

\def\fr#1#2{{{#1} \over {#2}}}

\def\ket#1{|{#1}\rangle}

\def\half{{\textstyle{1\over 2}}}
\def\frac#1#2{{\textstyle{{#1}\over {#2}}}}

\def\lsim{\mathrel{\rlap{\lower4pt\hbox{\hskip1pt$\sim$}}
    \raise1pt\hbox{$<$}}}
\def\gsim{\mathrel{\rlap{\lower4pt\hbox{\hskip1pt$\sim$}}
    \raise1pt\hbox{$>$}}}
\def\sqr#1#2{{\vcenter{\vbox{\hrule height.#2pt
	 \hbox{\vrule width.#2pt height#1pt \kern#1pt
	 \vrule width.#2pt}
	 \hrule height.#2pt}}}}

\begin{document}
 
\begin{flushright}
{IUHET 364\\}
{DF/IST-5.97\\}
{April 1997\\}
\end{flushright}
\vglue 1cm

\title{CPT, STRINGS, AND BARYOGENESIS}

\author{ O. BERTOLAMI }

\address{Departamento de F\'\i sica, Instituto Superior T\'ecnico,\\
Av. Rovisco Pais, Lisbon, Portugal}

\author{ D. COLLADAY, V.A. KOSTELECK\'Y }

\address{Department of Physics, Indiana University, Bloomington, IN 47405,
U.S.A.}

\author { R. POTTING \footnote{Speaker} }

\address{Sector de F\'\i sica, Universidade do Algarve, U.C.E.H., 8000 Faro,
Portugal}


\maketitle\abstracts{
In the context of string field theory,
the possibility exists for the
spontaneous violation of Lorentz invariance and CPT.
In this talk,
we review its status and some experimental constraints.
We discuss the possibility that stringy CPT violation could give rise
to a mechanism in which baryogenesis occurs in the early Universe
in thermal equilibrium and show that this can produce, under suitable
circumstances, a baryon asymmetry equal to the observed value.}

\section{Introduction}
One of the natural instances where particle physics meets cosmology
is in the problem of the generation of baryon asymmetry in the early
Universe.
Sakharov showed that the simultaneous violation of baryon number and
of C and CP symmetries, in the presence of nonequilibrium processes,
is sufficient for generating a baryon asymmetry.~\cite{sakharov}
Many specific mechanisms that implement these constraints have been
investigated.~\cite{yoshimura}$^{\!-\,}$\cite{kuzmin}

In this talk, we outline an alternative way to generate baryon number
in the early Universe, through the spontaneous violation of CPT 
symmetry.~\cite{bckp}
Spontaneous violation of CPT can in principle occur in certain string 
theories.~\cite{kp}
If CPT and baryon number are violated, baryon asymmetry can arise 
in thermal equilibrium.~\cite{dolgov}$^{\!-\,}$\cite{kaplan}
This mechanism would have the advantage of being 
otherwise independent of C- and CP- violating processes, which in a GUT
are typically contrived to match the observed baryon asymmetry and are
unrelated to the experimentally measured CP violation in the standard
model.

We describe here how CPT violation might occur spontaneously 
in string theory, 
and how it can give rise to baryogenesis along
the lines indicated above. We also discuss dilution effects through
electroweak sphalerons.
Other possible experimental implications of spontaneous CPT violation
are considered elsewhere.~\cite{kp2}$^{\!-\,}$\cite{Dtests}

\section{CPT violation in String Field Theory}
\subsection{Noncanonical vacua}
For definiteness, we use
Witten's version of (type I) string field theory:~\cite{witten}
\begin{equation}
L=\int\Phi*Q\Phi + g\int\Phi*\Phi*\Phi.
\end{equation}
Here,
$\Phi$ is the string field, which can be expanded in particle
modes:
\begin{eqnarray}
\ket\Phi &=&
\bigl[ \phi (x_0)
+ A_\mu (x_0) \al_{-1}^\mu
+{1\over\sqrt 2} i B_\mu (x_0) \alpha^\mu_{-2}\cr
&&+{1\over\sqrt 2} B_{\mu\nu}(x_0) \alpha_{-1}^\mu \alpha_{-1}^\nu 
+\beta_1 (x_0) b_{-1} c_{-1}
+\ldots \bigr]
\ket{-\half}.
\end{eqnarray}
First-quantized string theory is obtained as
perturbation theory around the solution $\Phi=0$ of the equation of
motion $Q\Phi +\Phi*\Phi=0$.~\cite{gmw}$^{\!-\,}$\cite{gm}
By no means, however, is this the only solution.
Other noncanonical solutions of the equations of motion for $\Phi$
exist.~\cite{ks}

Perturbation theory around those backgrounds yields different physics,
for instance:~\cite{ks,ks2,kp}
\begin{itemize}
\item
There are backgrounds with no tachyonic mode,
just as in the electroweak standard model
the would-be tachyon becomes a massive field
as the Higgs field takes a nonzero expectation value.
\item
Many of the originally (massive) degrees of freedom 
can become nonpropagating because string field theory
is nonlocal on the Planck scale, so derivatives in the
cubic term appear in the effective propagator for noncanonical
backgrounds.
\item
Lorentz covariance may be lost due to spontaneous symmetry breaking.
\item
Other discrete symmetries like CPT may be lost.
\end{itemize}

\subsection{CPT violation}
In string field theory, solutions exist in which scalar field
components have a nonzero value.
This in turn can lead to an effective action for tensor field components
that would give a vacuum expectation value to the latter.
As an example, consider the bosonic string field theory described above.
It contains the coupling $A_\mu A^\mu \phi$.
It follows that if $\phi$ gets a negative expectation value, this
would contribute in turn to a negative squared mass for $A_\mu$.
In that case, $A_\mu$ could get a vacuum expectation value,
breaking Lorentz invariance.
As $A_\mu$ is odd under CPT, this means that CPT would also be
broken.

A numerical investigation based on a level-cutoff scheme~\cite{expec}
has shown that nontrivial
solutions of bosonic string field theory do indeed exist.
Moreover, among those are solutions that break Lorentz invariance.
Evidence for CPT-violating solutions also appears. 

\section{Experimental constraints}

Consider the cubic coupling
\begin{equation}
T_{\mu_1...\mu_n}\bar\psi\Gamma^{\mu_1...\mu_r}
\partial^{\mu_{r+1}}...\partial^{\mu_n}\psi
\end{equation}
in which $T_{\mu_1...\mu_n}$ acquires a vacuum expectation value.
If $T$ is odd under CPT, this yields a CPT-violating chemical potential
$\mu$ for the fermion (quark) $\psi$.

The chemical potential can, for instance,
create an effective mass (or energy)
difference between particles and antiparticles, thus violating CPT.
The tightest experimental bounds on CPT violation
involve the neutral $K-\bar K$ system~\cite{gibbons}
\begin{equation}
{\Delta m\over m} < 2 \times 10^{-18}.
\end{equation}
For this reason, we expect CPT violation, 
if present, to be highly suppressed.
The natural suppression factor is the energy scale over the Planck mass.

An analysis~\cite{kp,bckp} shows that, for the case $k=0$,
any expectation value of the tensor field should
be suppressed by two powers of $m_l/M_{pl}$, and for the case $k=1$
at least by one. 
For $k\ge2$ the required suppression factors are automatically
provided by the derivatives.

\section{Baryogenesis}

\subsection{Baryogenesis in thermal equilibrium}
As indicated above, the presence of CPT violation gives rise to the
possibility of baryogenesis in the early Universe.
In the presence of processes violating baryon number (for instance,
at the GUT scale) in thermal equilibrium, a nonzero baryon number
is attained that is controlled by the value of the chemical
potential.

In this case, a calculation shows that
the contribution to the baryon-number asymmetry 
per comoving volume for $k=0$ is given by 
\begin{equation}
\fr{n_q - n_{\bar q}}{3s} \sim
{15g\over{2\pi^4 g_s(T)}}{\mu\over T}I_0(m_q/T)
\quad ,
\label{qcontrib}
\end{equation}
where $\mu$ is the chemical potential generated by the CPT-violating
term, and~\cite{bckp}
\begin{equation}
I_0(r)=\int_{r}^\infty
dx\,x\sqrt{x^2-r^2}e^x (1+e^x)^{-2}
\quad .
\end{equation}
The integral obeys the condition $I_0(r) < I_0(0) = \pi^2/6$.
If we take $\mu$ to be suppressed by two orders of $m_l/M_{pl}$,
or $\mu\sim m_l^2/M_{pl}$,
we find $n_B/s \sim (10^{-19} m_l /T) I_0(m_q/T)$,
far too small to reproduce the
observed value $n_B/s \simeq 10^{-10}$.

A similar analysis shows that the case $k=1$ also generates too
small a contribution to reproduce the observed baryon asymmetry.

However, $k=2$ generates a baryon asymmetry
\begin{equation}
{n_B\over s}\sim {3\over5}{T\over M_{pl}}.
\end{equation}
which, for appropriate decoupling temperature $T_D$, can reproduce the
observed value.
Note, however, that possible dilution mechanisms that might occur
subsequently must be taken into account.

\subsection{Dilution through sphaleron transitions}
It has been pointed out~\cite{kuzmin} that baryon asymmetry can be
diluted by the occurrence of sphaleron transitions, which
violate baryon number. These processes are expected to be unsuppressed
above the electroweak scale, and exceed the expansion rate of the
Universe below $10^{12}\,$GeV.~\cite{ak}

If the initial value of $B-L$ is zero,
an analysis~\cite{bckp} shows that 
the original asymmetry is diluted by a factor of about $10^{-6}$.
If initially $B-L\ne0$, the dilution by the
($B-L$ conserving) sphalerons is by a factor of order one.
In the former case, this means that the observed asymmetry is generated
if the initial baryon asymmetry takes place at a temperature of
$10^{-4}M_{pl}$, a value close to the GUT scale and the leptoquark
mass $M_X$, which is consistent with the requirement the
rate of baryon number violation exceeds the expansion rate of the 
Universe during that period.

\section{Summary}
We have explored the possibility that baryogenesis occurs through
spontaneous CPT breaking from string theory.
We have found that the CPT-breaking terms with $k=2$, accompanied by
interactions violating baryon number, generate a large baryon
asymmetry at the GUT scale.
If the interactions preserve $B-L$, the subsequent dilution through
sphaleron transition will then reproduce the observed baryon asymmetry.

\end{document}